\documentclass[prc,twocolumn,showpacs,superscriptaddress, twoside,showpacs]{revtex4} 

\usepackage{graphicx}
\usepackage{dcolumn}
\usepackage{color}
\usepackage{bm}
\usepackage{amsmath}
\usepackage{amsfonts}
\usepackage{amssymb}
\usepackage{threeparttable}
\usepackage{longtable}%
\allowdisplaybreaks[4]

\begin{document}

\title{New high-spin structure and possible chirality in $^{109}$In }

\author{M. Wang}
\affiliation{School of Physics and Nuclear Energy Engineering, Beihang University, Beijing 100191, China}
 
\author{Y.~Y. Wang}
\affiliation{School of Physics and Nuclear Energy Engineering, Beihang University, Beijing 100191, China}

\author{L.~H. Zhu}
\thanks{Corresponding author: zhulh@buaa.edu.cn}
\affiliation{School of Physics and Nuclear Energy Engineering, Beihang University, Beijing 100191, China}
\affiliation{Beijing Advanced Innovation Center for Big Date-based Precision Medicine, Beihang University, Beijing, 100083, China}

\author{B.~H. Sun}
\thanks{Corresponding author: bhsun@buaa.edu.cn}
\affiliation{School of Physics and Nuclear Energy Engineering, Beihang University, Beijing 100191, China}
\affiliation{Beijing Advanced Innovation Center for Big Date-based Precision Medicine, Beihang University, Beijing, 100083, China}

\author{G.~L. Zhang}
\affiliation{School of Physics and Nuclear Energy Engineering, Beihang University, Beijing 100191, China}
\affiliation{Beijing Advanced Innovation Center for Big Date-based Precision Medicine, Beihang University, Beijing, 100083, China}

\author{L.~C. He}
\affiliation{School of Physics and Nuclear Energy Engineering, Beihang University, Beijing 100191, China}

\author{W.~W. Qu}
\affiliation{School of Physics and Nuclear Energy Engineering, Beihang University, Beijing 100191, China}
\affiliation{State Key Laboratory of Radiation Medicine and Protection, School of Radiation Medicine and Protection, Soochow University, Suzhou 215123, China}

\author{F. Wang}
\affiliation{School of Physics and Nuclear Energy Engineering, Beihang University, Beijing 100191, China}

\author{T.~F. Wang}
\affiliation{School of Physics and Nuclear Energy Engineering, Beihang University, Beijing 100191, China}
\affiliation{Beijing Advanced Innovation Center for Big Date-based Precision Medicine, Beihang University, Beijing, 100083, China}

\author{Y.~Y. Chen}
\affiliation{School of Physics and Nuclear Energy Engineering, Beihang University, Beijing 100191, China}

\author{C. Xiong}
\affiliation{School of Physics and Nuclear Energy Engineering, Beihang University, Beijing 100191, China}

\author{J. Zhang}
\affiliation{School of Physics and Nuclear Energy Engineering, Beihang University, Beijing 100191, China}

\author{J.~M. Zhang}
\affiliation{School of Physics and Nuclear Energy Engineering, Beihang University, Beijing 100191, China}

\author{Y. Zheng}
\affiliation{China Institute of Atomic Energy, Beijing 102413, China}

\author{C.~Y. He}
\affiliation{China Institute of Atomic Energy, Beijing 102413, China}

\author{G.~S. Li}
\affiliation{China Institute of Atomic Energy, Beijing 102413, China}

\author{J.~L. Wang}
\affiliation{China Institute of Atomic Energy, Beijing 102413, China}

\author{X.~G. Wu}
\affiliation{China Institute of Atomic Energy, Beijing 102413, China}

\author{S.~H. Yao}
\affiliation{China Institute of Atomic Energy, Beijing 102413, China}

\author{C.~B. Li}
\affiliation{College of Physics, Jilin University, Changchun 130013, China}

\author{H.~W. Li}
\affiliation{College of Physics, Jilin University, Changchun 130013, China}

\author{S.~P. Hu}
\affiliation{College of Physics and Energy, Shenzhen University, Shenzhen 518060, China}

\author{J.~J. Liu}
\affiliation{College of Physics and Energy, Shenzhen University, Shenzhen 518060, China}

\date{\today}
\begin{abstract}

    High-spin structure of $^{109}$In has been investigated with the $^{100}$Mo($^{14}$N, 5$n$)$^{109}$In reaction at a beam energy of 78 MeV using the in-beam $\gamma$ spectroscopic method.
    The level scheme of $^{109}$In has been modified considerably and extended by 46 new $\gamma$-rays to the highest excited state at 8.979 MeV and $J^{\pi}$=(45/2$^{+}$).
    The new level scheme consists of eight bands, six of which are identified as dipole bands.
    The configurations have been tentatively assigned with the help of the systematics of neighboring odd-$A$ indium isotopes and the experimental aligned angular momenta. The dipole bands are then compared with the titled axis cranking calculation in the framework of covariant density function theory (TAC-CDFT).
    The results of theoretical calculation based on the configurations, which involve one proton hole at the $g_{9/2}$ orbital and two or four unpaired neutrons at $g_{7/2}$, $d_{5/2}$ and $h_{11/2}$ orbitals, show that the shape of $^{109}$In undergoes an evolution on both $\beta$ and $\gamma$ deformations and possible chirality is suggested in $^{109}$In.

\end{abstract}
\pacs{21.10.Re, 23.20.Lv, 27.60.+j} \maketitle
\date{today}

\section{Introduction}
Tremendous work has been dedicated to study the nuclei in the $A\sim110$ mass region.
For nucleus in this region, the $d_{5/2}$, $g_{7/2}$, high-$j$ $h_{11/2}$-intruder neutron orbitals and the high-$\Omega$ $g_{9/2}$ proton orbital are lying near the Fermi surface, thus exhibiting many interesting phenomena, such as the signature inversion~\cite{RhAg,SI_108Ag}, superdeformation~\cite{SF_104105Pb,SF_108Cd}, magnetic rotation~\cite{M1_106Ag,M1_107Ag,MR112In}, chiral rotation~\cite{chiral_106Rh,chiral_105Rh,chiral_104Rh,chiral_103Rh,chiral106Ag}, shape coexistence~\cite{shape98Mo,Shapecoexistence106Pd,Shapecoexistence106Ag}, smooth band termination~\cite{sbt109Sb,sbt108Sn} and giant dipole resonance built on highly excited states~\cite{Giantresonances111Sn,GiantresonancesSn}.

Deformation due to the proton particle excitation across the $Z=50$ shell gap into one of $d_{5/2}$, $g_{7/2}$ and $h_{11/2}$ orbitals, increases the overlap of valence neutron and proton.
The Fermi levels for the protons and the neutrons of the indium isotopes are in different intruder orbitals.
The orbital near the top of the $\pi$$g_{9/2}$ is oblate driving, whereas the orbital near the bottom of the $\nu$$h_{11/2}$ tends to drive the nucleus to a prolate shape, which makes indium isotopes exhibiting various deformations.
Furthermore, some bands in Rh and Ag isotopes with $A\sim110$ show characteristics of the nuclear chirality~\cite{chiral_100Tc,chiral_103Rh,chiral_104Rh,chiral_105Rh,chiral_106Rh,chiral_106Ag}, which take places when the angular momenta of the valence proton, the valence neutron and the core rotation tend to be mutually perpendicular \cite{SFMeng}.
However, chirality in the indium isotopes has not been reported.

The excited states in $^{109}$In were first identified by Poelgeest~\emph{et al.}~\cite{Van1979} using the $^{107}$Ag($a$, $2n$)$^{109}$In reaction, and an isomeric state of $T_{1/2}$=210 ms was observed.
Since then $^{109}$In was investigated successively by Kownacki~\emph{et al.}~\cite{Kownacki1997} using the $^{92}$Mo($^{19}$F, 2$p$)$^{109}$In reaction and Negi~\emph{et al}.~\cite{Negi2012} using the $^{96}$Zr($^{19}$F, 6$n $)$^{109}$In reaction, in which a level scheme up to $J$=35/2 at 5.89 MeV was established and the shears mechanism in $^{109}$In was discussed within the framework of the titled axis cranking (TAC).
Furthermore, for the neighboring isotopes and isotones of $^{109}$In, a lot of interesting nuclear phenomena have been observed, such as magnetic rotation in $^{107-108,110-115}$In~\cite{MR_107In,MR_108110In,MR_111In,MR_112In,MR_113In,113In,MR_114In,MR_115In}, $^{108}$Cd~\cite{Shearsband108Cd} and $^{107}$Ag~\cite{M1_107Ag}, antimagnetic rotation in $^{108}$Cd~\cite{antimr_108Cd} and $^{112}$In~\cite{antiMR_112In}, chiral bands in $^{107}$Ag~\cite{chiral_107Ag} and $^{105}$Rh~\cite{chiral_105Rh}.

\begin{figure*}
  \includegraphics[width=17cm]{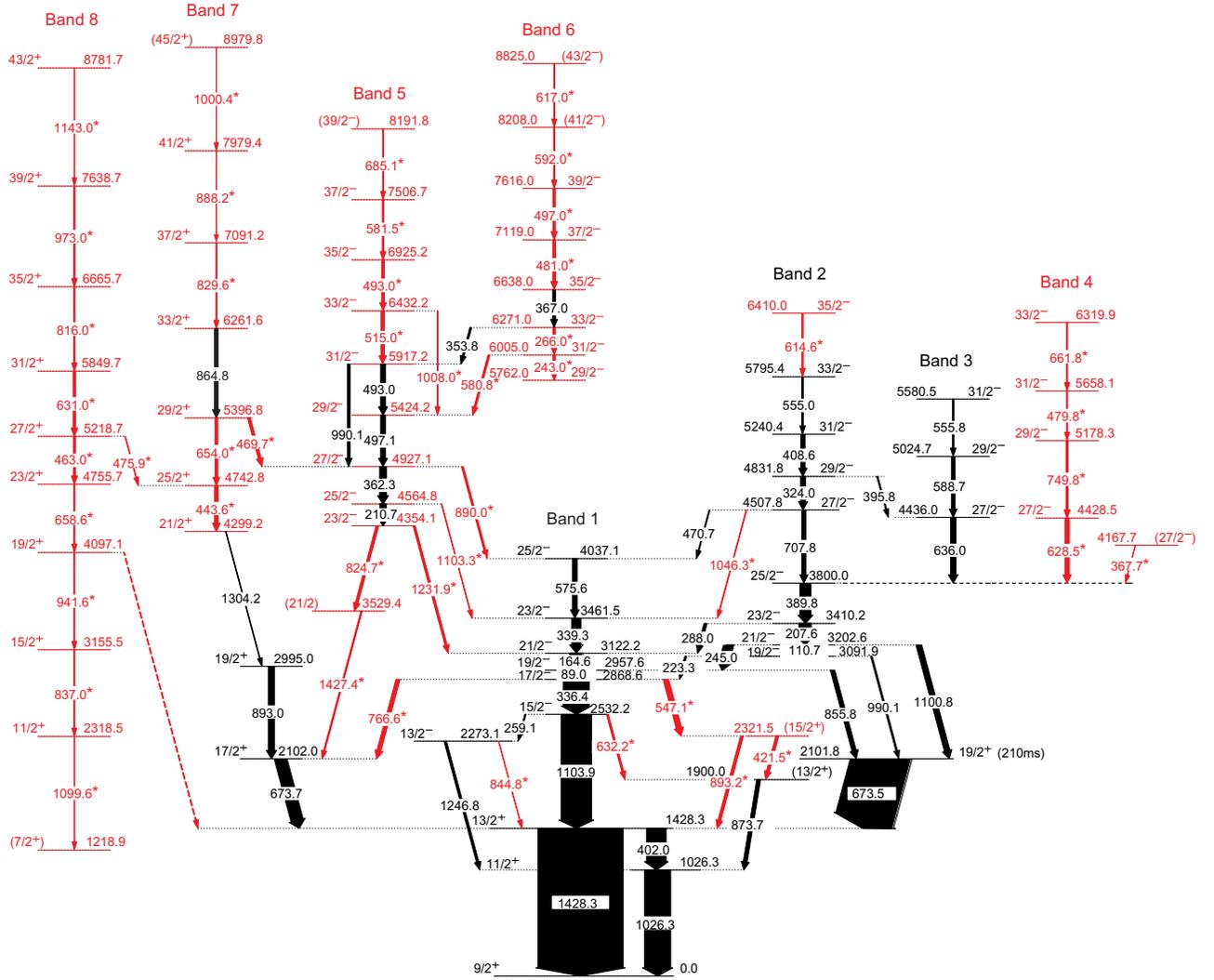}
  \caption{\label{FIG:109In} (Color online) The level scheme of $^{109}$In proposed in the present work. The scheme in red shows the newly identified part and the transitions marked with an asterisk($^{*}$) are newfound.}
\end{figure*}

In this paper, we present new experimental results and theoretical studies on the high-spin structures in $^{109}$In.
The level scheme has been significantly modified and 5 new bands are identified. The configurations of bands in $^{109}$In have been analyzed with the help of the systematics of neighboring odd-$A$ indium isotopes and the experimental aligned angular momenta. The tilted axis cranking covariant density function theory (TAC-CDFT) is used to investigate the intrinsic structure of $^{109}$In.
The $\gamma$ deformation, shape evolution and the possible chirality will be discussed.


\section{EXPERIMENTAL DETAILS}\label{sec:details}
Excited states in $^{109}$In were populated using the $^{100}$Mo($^{14}$N, 5$n$)$^{109}$In fusion-evaporation reaction at a beam energy of 78 MeV. The beam was delivered by the HI-13 tandem accelerator at the China Institute of Atomic Energy (CIAE) in Beijing.
The target consists of a 0.5 mg/cm$^{2}$ foil of $^{100}$Mo with a backing of 10 mg/cm$^{2}$ thick $^{197}$Au .
The $\gamma$-rays were detected by an array composed of 9 BGO-Compton-suppressed HPGe detectors, two low-energy photon (LEP) HPGe detectors and one clover detector. Three HPGe detectors and the clover detector were placed at around 90$^{\circ}$, six HPGe detectors at around ±40$^{\circ}$ and two LEP HPGe detectors at around ±60$^{\circ}$ relative to the beam direction. The energy and efficiency calibrations of the detectors were performed using the standard sources of $^{152}$Eu and $^{133}$Ba. A total of 84$\times$10$^{6}$ $\gamma-\gamma$ coincidence events were recorded in an event-by-event mode. The data have been sorted into a fully symmetrized $E_{\gamma}$-$E_{\gamma}$ matrix and analyzed using the software package RADWARE~\cite{Radford} for the $\gamma$-ray coincidence relationship.

To obtain information on the multipolarities of $\gamma$ rays, the ratios of directional correlation of oriented states (DCO) have been analyzed from an asymmetric DCO matrix obtained by sorting the data from the detectors at $\sim\pm$40$^{\circ}$ on one axis and the data from the detectors at $\sim$90$^{\circ}$ on the other axis.

\begin{figure*}
  \includegraphics[width=17.5cm]{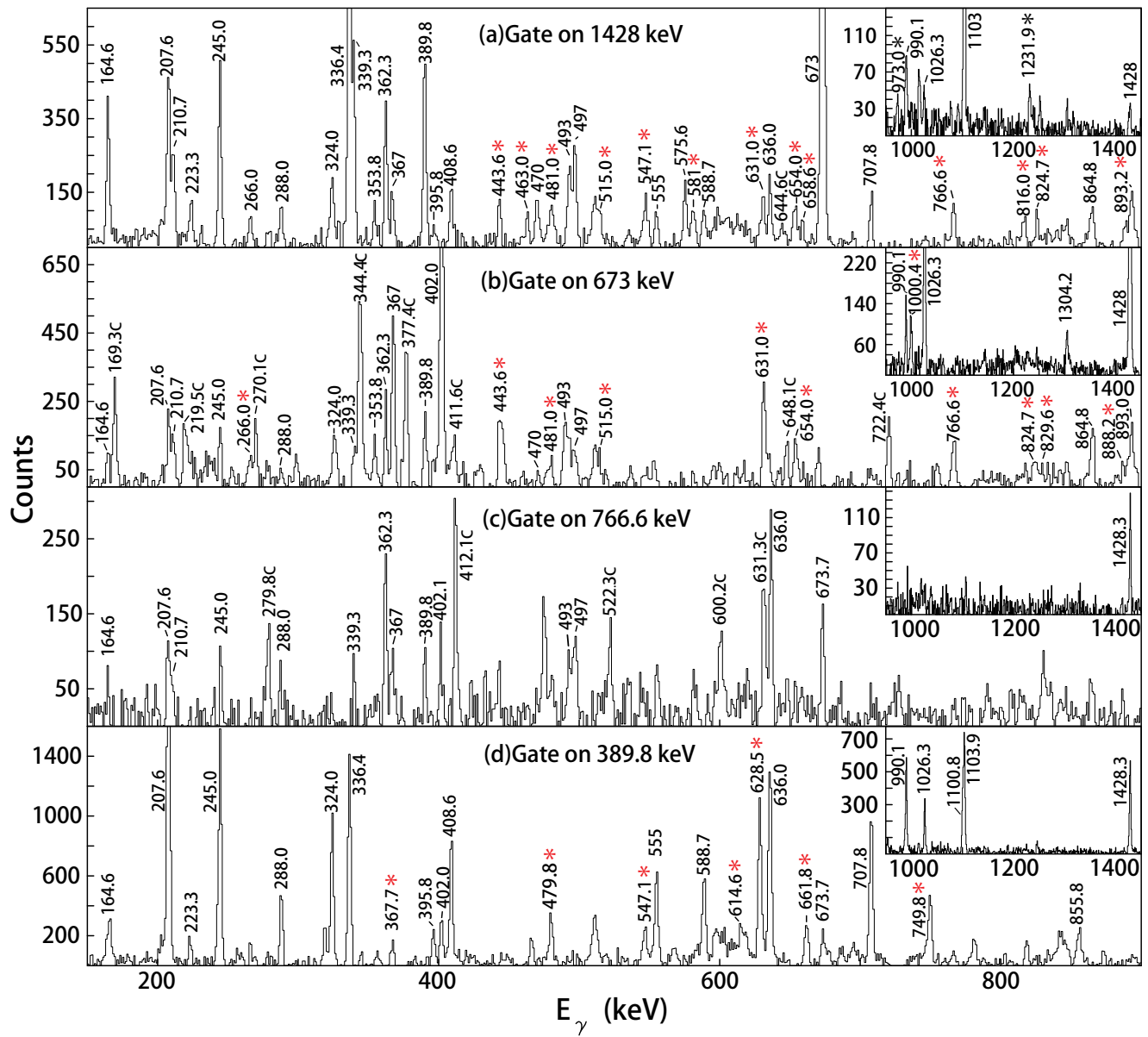}
  \caption{\label{FIG:1428} (Color online) $\gamma$-ray coincidence spectra with gates set on the (a) 1428-keV, (b) 673-keV, (c) 766.6-keV and (d) 389.8-keV transitions. Insets show the higher energy part of the spectra. The energies marked by the asterisks are newly identified $\gamma$-rays and the energies marked by C are contaminants.}
\end{figure*}

\begin{figure*}
  \includegraphics[width=17.5cm]{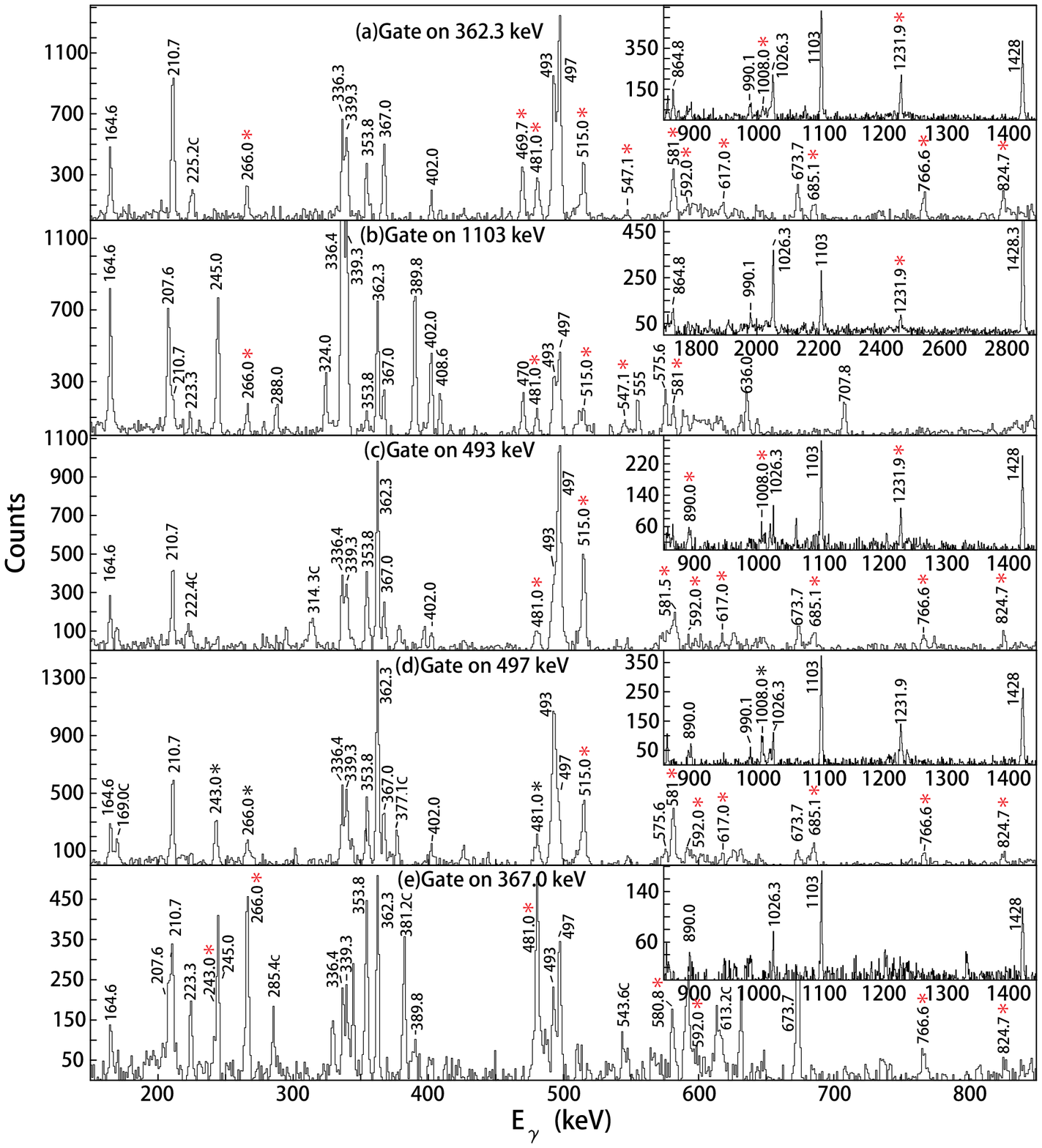}
  \caption{\label{FIG:1103} (Color online) $\gamma$-ray coincidence spectra with gates set on (a) 362.3-keV, (b) 1103-keV, (c) 493-keV, (d) 497-keV and (e) 367.0-keV transitions. Insets show the higher energy part of the spectra. The energies marked by the asterisks are newly identified $\gamma$-rays and the energies marked by C are contaminants.}
\end{figure*}

\section{EXPERIMENTAL RESULTS}\label{sec:experimentalresults}

The level scheme of $^{109}$In deduced from the present work is shown in Fig.~\ref{FIG:109In}.
The $\gamma$-ray energies, intensities, DCO ratios and spin-parity assignments of levels in $^{109}$In extracted from present experimental data are listed in Table.~\ref{Table:exp}.
The placements of $\gamma$-rays in the level scheme are determined through the observed coincidence relationships, intensity balances, and energy sums. The levels have been grouped into 8 bands. Compared with the results reported in Ref.~\cite{Negi2012}, the level scheme of $^{109}$In has been significantly extended and revised. 46 new $\gamma$-rays and 5 new band structures have been added.
In the spectrum gated on the 1428-keV transition, as shown in Fig.~\ref{FIG:1428}(a), one can see most of the corresponding coincidence $\gamma$-peaks of $^{109}$In.

\subsection{Coincidence relationships of the prompt 673-keV transition}\label{sec:band123}

The near degeneracy of the 17/2$^{+}$ with excitation energy of 2102.0 keV and the 19/2$^{+}$ states with excitation energy of 2101.8 keV particularly deserves justification.
A doublet $\gamma$-ray of 673 keV was reported in the earlier works~\cite{Negi2012,Kownacki1997,Van1979}.
It was suggested that the prompt 673-keV transition is originated from 17/2$^{+}$ level to 13/2$^{+}$ level~ \cite{Kownacki1997,Negi2012} and the delayed 673-keV $\gamma$-ray decays from the 210 ms isomeric state 19/2$^{+}$ to the yrast 13/2$^{+}$ state of 1428.3 keV~\cite{Negi2012,Van1979}.
The multipolarities of the delayed and the prompt 673-keV transitions were confirmed by the angular distribution measurement and the conversion electron measurement as the $M3$ and $E2$ respectively in Ref.~\cite{Van1979}.
The existence of the doublet 673-keV transitions is certain. However, we modified and supplemented the levels according to the coincidences between 673-keV $\gamma$-rays and the transitions from the higher levels.

In Fig.~\ref{FIG:1428}(b), the spectrum gated on the 673-keV transition, the $\gamma$-rays from higher levels, such as the 164.6-, 339.3-keV transitions of band 1 and the 207.6-, 389.8-keV transitions of band 2 can be clearly seen.
On account of the 210 ms isomeric level, the 673-keV transition should not be in coincidence with $\gamma$-rays which feed from the levels above this isomeric level.
Therefore, those $\gamma$-rays depopulated from higher levels are coincident with the prompt 673-keV transition, which exact energy is 673.7 keV.
Moreover, the 990.1-keV $\gamma$-ray was found to be coincident with the 673.7-keV transition.
In Refs.~\cite{Kownacki1997,Negi2012}, the doublet 990-keV transition was taken as the decay path for the levels up the 19/2$^{-}$ state at 3091.9 keV to the prompt 17/2$^{+}$ state at 2102.0 keV.
However, the coincidence of the 673.7-keV transition with the 339.3- and 164.6-keV transitions could not be explained.
In our work, the newly added 1427.4- and 766.6-keV transitions, which feed the prompt level at 2102.0 keV, give reasonable explanations to the peculiar coincidence of the 673.7-keV transition and the doublet $\gamma$-ray of 990.1 keV is placed in band 5, which will be explained in Section~\ref{sec:band56}.

The $\gamma$-ray of 766.6 keV was previously reported as a transition feeding the yrast 13/2$^{+}$ state at 1428.3 keV in Ref.~\cite{Van1979}.
From Fig.~\ref{FIG:1428}(c), the spectrum gated on the 766.6-keV transition, $\gamma$-rays of 673 keV and all the transitions depopulated from the levels above the 17/2$^{-}$ level at 2868.6 keV, such as the 164.6-, 339.3-, 245.0- and 389.8-keV transitions, can be identified.
Meanwhile, the 766.6-keV transition has no coincidence with the 336.4-keV transition and the $\gamma$-rays from lower excited states. Therefore, the 766.6-keV transition is placed parallel to the 336.4-keV transition and feeds to the 17/2$^{+}$ state at 2102.0 keV.

From the spectrum gated on the 1428-keV transition, as shown in Fig.~\ref{FIG:1428}(a), a doublet 1428-keV transitions (the exact energies are 1427.4 and 1428.3 keV, respectively) have been observed. Moreover, the 1427.4-keV transition has mutual coincidence with the 824.7-keV transition. The coincidence relationship and the energy balances confirm the placement of 1427.4-keV transition.
The 1427.4- and 766.6-keV transitions are in the decay path for those upper levels of bands 1, 5 and 6 to the prompt level at 2102.0 keV. These transitions help to pin down the prompt 673-keV transition depopulated from the 17/2$^{+}$ state at 2102-keV level.



\subsection{Bands 1, 2, 3 and 4}\label{sec:band1234}

Bands 1-4 consist of a series of $M1$ transitions.
From the spectra gated on the 1428-keV and the 389.8-keV transitions, shown in Figs.~\ref{FIG:1428}(a) and (d), most of the cascades in those bands can be identified.
The sequence of those $\gamma$-rays along with the 288.0-, 245.0- and 223.3-keV transitions suggested in Ref.~\cite{Negi2012} has been verified in the present work.
The newly added 614.6-keV transition of band 2 has mutual coincidence with all the $\gamma$-rays of this band and the lower part of band 1.
We thus place this $\gamma$-ray on the top of band 2 based on intensities.
Band 2 is associated with band 3 by $\gamma$-transitions with the energy of 395.8 keV.
Besides, band 2 also feeds to band 1 via the 470.7-keV and the 1046.3-keV transitions.
The $\gamma$-ray of 1046.3 keV is a new transition observed in the present work. It has coincidence with $\gamma$-rays of band 2 and the 339.3-, 164.6-, 336.4-keV transitions, etc.
However, it has no coincidence with the 575.6-keV transition. Therefore, this $\gamma$-ray is placed parallel to the 470.7-keV transition.

Band 4, a newly established band in this work, is composed of a sequence of $\triangle$$I$=1 transitions.
The $\gamma$-rays of this band have mutual coincidences with each other and the transitions which depopulate the states below 25/2$^{-}$ level at 3800.0 keV of band 2.
From the 389.8-keV gated spectrum, as shown in Fig.~\ref{FIG:1428}(d), all the members of this band can be identified.


\setlength{\tabcolsep}{18pt}
\begin{table*}
\centering
\caption{\label{Table:exp}The $\gamma$-ray energies, intensities, DCO ratios, and the initial and final state spin-parities of $^{109}$In deduced in the present work.}
\begin{tabular}{cccccc}
\hline
\hline
$E_{\gamma}$(keV)$^{a}$  &$E_{i}$$\rightarrow$$E_{f}$$^{b}$  &$I_{\gamma}$($\%$)$^{c}$ &$R_{DCO}$(D)$^{d}$ &  $R_{DCO}$(Q)$^{e}$  &$I_{i}^{\pi}$$\rightarrow$$I_{f}$$^{\pi}$$^{f}$\\
\hline
89.0&2957.6$\rightarrow$2868.6&10(1)& & &19/2$^{-}$$\rightarrow$17/2$^{-}$ \\
110.7&3202.6$\rightarrow$3091.9 &7(2)&&&21/2$^{-}$$\rightarrow$19/2$^{-}$\\
164.6&3122.2$\rightarrow$2957.6 &38(2)&&&21/2$^{-}$$\rightarrow$19/2$^{-}$\\
207.6&3410.2$\rightarrow$3202.6 &38(2)&&&23/2$^{-}$$\rightarrow$21/2$^{-}$\\
210.7&4564.8$\rightarrow$4354.1 &20(1)&&&25/2$^{-}$$\rightarrow$23/2$^{-}$\\
223.3&3091.9$\rightarrow$2868.6 &5.6(4)&1.18(30)&&19/2$^{-}$$\rightarrow$17/2$^{-}$\\
243.0&6005.0$\rightarrow$5762.0 &7(1)&1.02(10)&&31/2$^{-}$$\rightarrow$29/2$^{-}$\\
245.0&3202.6$\rightarrow$2957.6 &31(1)&1.30(13)&&21/2$^{-}$$\rightarrow$19/2$^{-}$\\
259.1&2532.2$\rightarrow$2273.1 &3.9(6)&&&15/2$^{-}$$\rightarrow$13/2$^{-}$\\
266.0&6271.0$\rightarrow$6005.0 &6.8(5)&1.17(13)&&33/2$^{-}$$\rightarrow$31/2$^{-}$\\
288.0&3410.2$\rightarrow$3122.2 &9.2(6)&0.82(16)&&23/2$^{-}$$\rightarrow$21/2$^{-}$\\
324.0&4831.8$\rightarrow$4507.8 &16.2(9)&1.32(43)&&29/2$^{-}$$\rightarrow$27/2$^{-}$\\
336.4&2868.6$\rightarrow$2532.2 &82(3)&0.79(5)&&17/2$^{-}$$\rightarrow$15/2$^{-}$\\
339.3&3461.5$\rightarrow$3122.2 &29(1)&0.61(5)&&23/2$^{-}$$\rightarrow$21/2$^{-}$\\
353.8&6271.0$\rightarrow$5917.2 &5.8(4)&1.05(13)&&33/2$^{-}$$\rightarrow$31/2$^{-}$\\
362.3&4927.1$\rightarrow$4564.8 &22(2)&1.05(14)&&27/2$^{-}$$\rightarrow$25/2$^{-}$\\
367.0&6638.0$\rightarrow$6271.0 &9(1)&0.99(14)&&35/2$^{-}$$\rightarrow$33/2$^{-}$\\
367.7&4167.7$\rightarrow$3800.0 &1(1)&&&(27/2$^{-}$)$\rightarrow$25/2$^{-}$\\
389.8&3800.0$\rightarrow$3410.2 &35(2)&1.06(15)&&25/2$^{-}$$\rightarrow$23/2$^{-}$\\
395.8&4831.8$\rightarrow$4436.0 &3.8(3)&&&29/2$^{-}$$\rightarrow$27/2$^{-}$\\
402.0&1428.3$\rightarrow$1026.3 &62(3)&0.61(17)&&13/2$^{+}$$\rightarrow$11/2$^{+}$\\
408.6&5240.4$\rightarrow$4831.8 &13.6(8)&1.14(11)&&31/2$^{-}$$\rightarrow$29/2$^{-}$\\
421.5&2321.5$\rightarrow$1900.0 &9(3)&1.28(44)&&(15/2$^{+}$)$\rightarrow$(13/2$^{+}$)\\
443.6&4742.8$\rightarrow$4299.2 &10.9(7)&&0.82(17)&25/2$^{+}$$\rightarrow$21/2$^{+}$\\
463.0&5218.7$\rightarrow$4755.7 &7.2(5)&&1.07(16)&27/2$^{+}$$\rightarrow$23/2$^{+}$\\
469.7&5396.8$\rightarrow$4927.1 &9(3)&0.64(8)&&29/2$^{+}$$\rightarrow$27/2$^{-}$\\
470.7&4507.8$\rightarrow$4037.1 &3(2)&1.20(31)&&27/2$^{-}$$\rightarrow$25/2$^{-}$\\
475.9&5218.7$\rightarrow$4742.8 &2(1)&&0.60(10)&27/2$^{+}$$\rightarrow$25/2$^{+}$\\
479.8&5658.1$\rightarrow$5178.3 &3(1)&0.81(17)&&31/2$^{-}$$\rightarrow$29/2$^{-}$\\
481.0&7119.0$\rightarrow$6638.0 &9(1)&0.85(7)&&37/2$^{-}$$\rightarrow$35/2$^{-}$\\
493.0&5917.2$\rightarrow$5424.2 &16.1(8)&&&31/2$^{-}$$\rightarrow$29/2$^{-}$\\
493.0&6925.2$\rightarrow$6432.2 &8(1)&&&35/2$^{-}$$\rightarrow$33/2$^{-}$\\
497.0&7616.0$\rightarrow$7119.0 &7(1)&&&39/2$^{-}$$\rightarrow$37/2$^{-}$\\
497.1&5424.2$\rightarrow$4927.1 &16(2)&&&29/2$^{-}$$\rightarrow$27/2$^{-}$\\
515.0&6432.2$\rightarrow$5917.2 &10.3(7)&0.54(8)&&33/2$^{-}$$\rightarrow$31/2$^{-}$\\
547.1&2868.6$\rightarrow$2321.5 &17.5(10)&0.90(17)&&17/2$^{-}$$\rightarrow$(15/2$^{+}$)\\
555.0&5795.4$\rightarrow$5240.4 &4.7(4)&1.10(14)&&33/2$^{-}$$\rightarrow$31/2$^{-}$\\
555.8&5580.5$\rightarrow$5024.7 &4.8(8)&0.85(20)&&31/2$^{-}$$\rightarrow$29/2$^{-}$\\
575.6&4037.1$\rightarrow$3461.5 &15.1(9)&0.90(7)&&25/2$^{-}$$\rightarrow$23/2$^{-}$\\
580.8&6005.0$\rightarrow$5424.2 &6(1)&0.81(7)&&31/2$^{-}$$\rightarrow$29/2$^{-}$\\
581.5&7506.7$\rightarrow$6925.2 &5.0(10)&1.32(15)&&37/2$^{-}$$\rightarrow$35/2$^{-}$\\
588.7&5024.7$\rightarrow$4436.0 &11.6(7)&0.95(14)&&29/2$^{-}$$\rightarrow$27/2$^{-}$\\
592.0&8208.0$\rightarrow$7616.0 &4.4(4)&&&(41/2$^{-}$)$\rightarrow$39/2$^{-}$\\
614.6&6410.0$\rightarrow$5795.4 &4.0(4)&0.58(10)&&35/2$^{-}$$\rightarrow$33/2$^{-}$\\
617.0&8825.0$\rightarrow$8208.0 &3(1)&&&(43/2$^{-}$)$\rightarrow$(41/2$^{-}$)\\
628.5&4428.5$\rightarrow$3800.0 &13.9(9)&0.70(7)&&27/2$^{-}$$\rightarrow$25/2$^{-}$\\
631.0&5849.7$\rightarrow$5218.7 &7.8(9)&&1.04(11)&31/2$^{+}$$\rightarrow$27/2$^{+}$\\
632.2&2532.2$\rightarrow$1900.0 &6.0(9)&&&15/2$^{-}$$\rightarrow$(13/2$^{+}$)\\
636.0&4436.0$\rightarrow$3800.0 &17.5(10)&0.69(5)&&27/2$^{-}$$\rightarrow$25/2$^{-}$\\
654.0&5396.8$\rightarrow$4742.8 &8.6(6)&&0.83(10)&29/2$^{+}$$\rightarrow$25/2$^{+}$\\
658.6&4755.7$\rightarrow$4097.1 &3.2(3)&&0.93(9)&23/2$^{+}$$\rightarrow$19/2$^{+}$\\
661.8&6319.9$\rightarrow$5658.1 &3.2(5)&0.59(32)&&33/2$^{-}$$\rightarrow$31/2$^{-}$\\
673.5&2101.8$\rightarrow$1428.3 &188(7)&&&19/2$^{+}$$\rightarrow$13/2$^{+}$\\
673.7&2102.0$\rightarrow$1428.3 &40(2)&1.62(23)&&17/2$^{+}$$\rightarrow$13/2$^{+}$\\
685.1&8191.8$\rightarrow$7506.7 &2.2(3)&0.71(33)&&(39/2$^{-}$)$\rightarrow$37/2$^{-}$\\
707.8&4507.8$\rightarrow$3800.0 &12.8(8)&0.95(8)&&27/2$^{-}$$\rightarrow$25/2$^{-}$\\
749.8&5178.3$\rightarrow$4428.5 &6.5(5)&0.50(9)&&29/2$^{-}$$\rightarrow$27/2$^{-}$\\
766.6&2868.6$\rightarrow$2102.0 &13.0(8)&&1.45(35)&17/2$^{-}$$\rightarrow$17/2$^{+}$\\
816.0&6665.7$\rightarrow$5849.7 &4.1(4)&&1.19(16)&35/2$^{+}$$\rightarrow$31/2$^{+}$\\
\hline
\hline
\end{tabular}
\end{table*}

\setcounter{table}{0}
\begin{table*}
\caption{(Continued)}
\centering
\begin{tabular}{cccccc}
\hline
\hline
$E_{\gamma}$(keV)$^{a}$  &$E_{i}$$\rightarrow$$E_{f}$$^{b}$  &$I_{\gamma}$($\%$)$^{c}$ &$R_{DCO}$(D)$^{d}$ &  $R_{DCO}$(Q)$^{e}$  &$I_{i}^{\pi}$$\rightarrow$$I_{f}$$^{\pi}$$^{f}$\\
\hline
824.7&4354.1$\rightarrow$3529.4 &8.5(6)&0.84(22)&&23/2$^{-}$$\rightarrow$(21/2)\\
829.6&7091.2$\rightarrow$6261.6 &3.2(5)&&0.96(29)&37/2$^{+}$$\rightarrow$33/2$^{+}$\\
837.0&3155.5$\rightarrow$2318.5 &2.7(7)&&0.93(7)&15/2$^{+}$$\rightarrow$11/2$^{+}$\\
844.8&2273.1$\rightarrow$1428.3 &1(1)&&&13/2$^{-}$$\rightarrow$13/2$^{+}$\\
855.8&2957.6$\rightarrow$2101.8 &14(2)&&&19/2$^{-}$$\rightarrow$19/2$^{+}$\\
864.8&6261.6$\rightarrow$5396.8 &11.6(7)&&0.80(7)&33/2$^{+}$$\rightarrow$29/2$^{+}$\\
873.7&1900.0$\rightarrow$1026.3 &12.5(8)&1.16(36)&&(13/2$^{+}$)$\rightarrow$11/2$^{+}$\\
888.2&7979.4$\rightarrow$7091.2 &4.3(6)&&0.78(17)&41/2$^{+}$$\rightarrow$37/2$^{+}$\\
890.0&4927.1$\rightarrow$4037.1 &5.9(5)&&&27/2$^{-}$$\rightarrow$25/2$^{-}$\\
893.0&2995.0$\rightarrow$2102.0 &20(2)&&0.46(8)&19/2$^{+}$$\rightarrow$17/2$^{+}$\\
893.2&2321.5$\rightarrow$1428.3 &8(3)&&&(15/2$^{+}$)$\rightarrow$13/2$^{+}$\\
941.6&4097.1$\rightarrow$3155.5 &2.6(3)&&1.10(13)&19/2$^{+}$$\rightarrow$15/2$^{+}$\\
973.0&7638.7$\rightarrow$6665.7 &4.5(5)&&0.98(13)&39/2$^{+}$$\rightarrow$35/2$^{+}$\\
990.1&5917.2$\rightarrow$4927.1 &8.6(7)&1.30(41)&&31/2$^{-}$$\rightarrow$27/2$^{-}$\\
990.1&3091.9$\rightarrow$2101.8 &5.0(5)&&&19/2$^{-}$$\rightarrow$19/2$^{+}$\\
1000.4&8979.8$\rightarrow$7979.4 &0.4(3)&&&(45/2$^{+}$)$\rightarrow$41/2$^{+}$\\
1008.0&6432.2$\rightarrow$5424.2 &2.6(3)&&&33/2$^{-}$$\rightarrow$29/2$^{-}$\\
1026.3&1026.3$\rightarrow$0 &84(3)&0.59(7)&&11/2$^{+}$$\rightarrow$9/2$^{+}$\\
1046.3&4507.8$\rightarrow$3461.5 &2.4(3)&&&27/2$^{-}$$\rightarrow$23/2$^{-}$\\
1099.6&2318.5$\rightarrow$1218.9 &1.3(2)&&&11/2$^{+}$$\rightarrow$(7/2$^{+}$)\\
1100.8&3202.6$\rightarrow$2101.8 &18(1)&&&21/2$^{-}$$\rightarrow$19/2$^{+}$\\
1103.3&4564.8$\rightarrow$3461.5 &2(2) &&&25/2$^{-}$$\rightarrow$23/2$^{-}$\\
1103.9&2532.2$\rightarrow$1428.3 &98(3) &&&15/2$^{-}$$\rightarrow$13/2$^{+}$\\
1143.0&8781.7$\rightarrow$7638.7 &1.2(6)&&1.00(26)&43/2$^{+}$$\rightarrow$39/2$^{+}$\\
1231.9&4354.1$\rightarrow$3122.2 &6.8(5)&0.51(12)&&23/2$^{-}$$\rightarrow$21/2$^{-}$\\
1246.8&2273.1$\rightarrow$1026.3 &7.7(6)&0.57(25)&&13/2$^{-}$$\rightarrow$11/2$^{+}$\\
1304.2&4299.2$\rightarrow$2995.0 &2.1(3)&&0.54(11)&21/2$^{+}$$\rightarrow$19/2$^{+}$\\
1427.4&3529.4$\rightarrow$2102.0 &4.1(5)&&&(21/2)$\rightarrow$17/2$^{+}$\\
1428.3&1428.3$\rightarrow$0 &273(16)&&&13/2$^{+}$$\rightarrow$9/2$^{+}$\\
\hline
\hline
\end{tabular}
\begin{threeparttable}
\begin{tablenotes}
\item[a]Uncertainties are between 0.2 and 0.5 keV depending upon their intensity.
\item[b]Excitation energies of initial and final states, $E_{i}$ and $E_{f}$.
\item[c]Intensities are normalized to the 1103-keV transition with $I_{\gamma}=100$.
\item[d]DCO ratios gated by dipole transitions.
\item[e]DCO ratios gated by quadrupole transitions.
\item[f]Proposed spin and parity assignments to the initial $J^{\pi}_{i}$ and final $J^{\pi}_{f}$ levels.
\end{tablenotes}
\end{threeparttable}
\end{table*}

\subsection{Bands 5 and 6}\label{sec:band56}

\begin{figure*}
  \includegraphics[width=18cm]{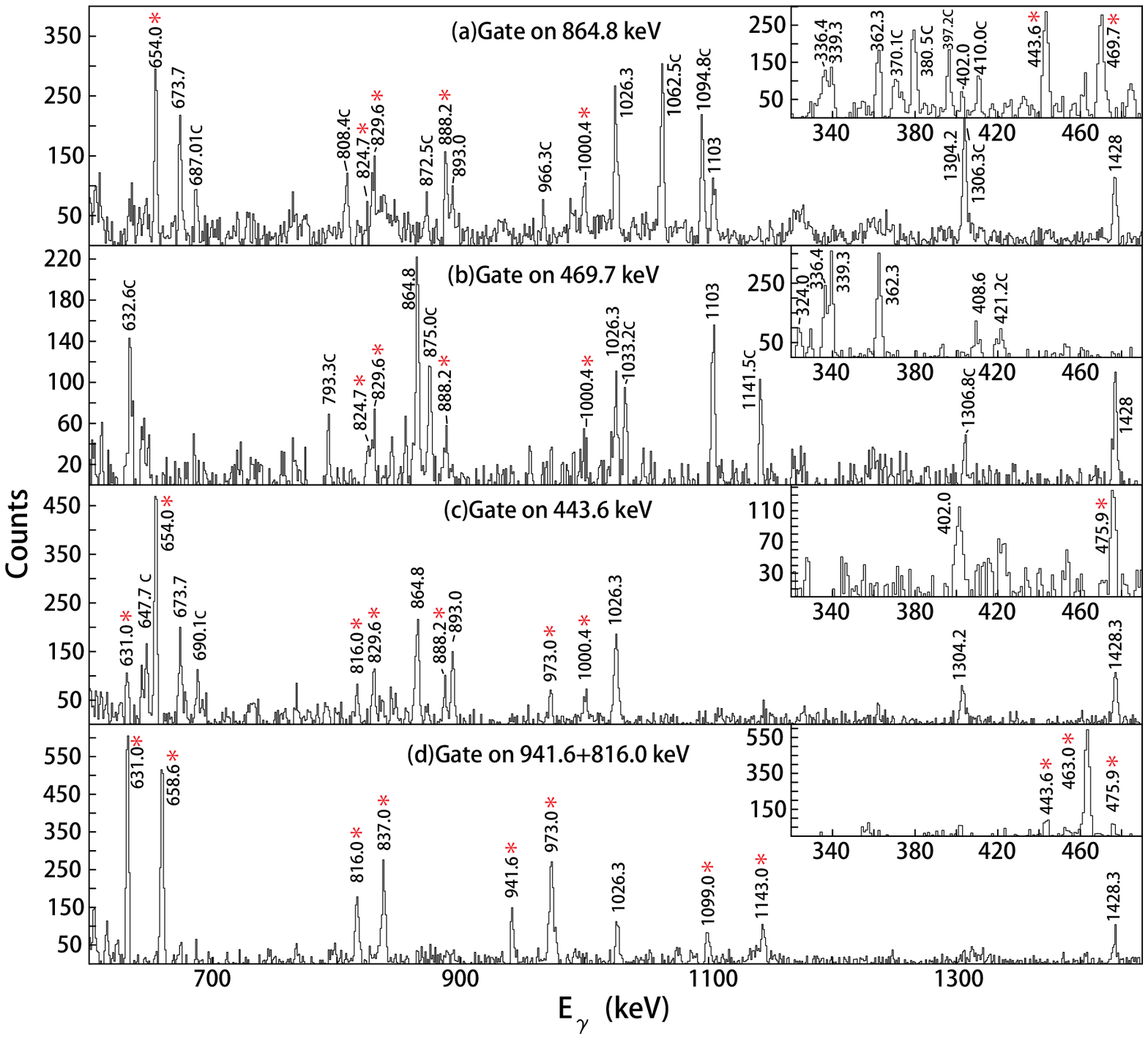}
  \caption{\label{FIG:443} (Color online) $\gamma$-ray coincidence apectra with gates set on (a) 864.8-keV, (b) 469.7-keV, (c) 443.6-keV and (d) 941.6+816.0-keV. Insets show the lower energy part of the spectra. The energies marked by the asterisks are newly identified $\gamma$-rays and the energies marked by C are contaminants.}
\end{figure*}

Bands 5 and 6 are well developed dipole bands. Parts of the transitions in those bands have been reported in Ref.~\cite{Negi2012}.
However, we modified and rearranged those $\gamma$-rays according to the newly found transitions and coincidence relationship.

The 23/2$^{-}$ level at 4354.1 keV is assigned to the bandhead of band 5.
The newly found transition with energy of 210.7 keV has mutual coincidence with all the members of band 5 and the transitions with energies of 164.6, 336.4, 1428 keV, etc.
Moreover, it has coincidence with the 1231.9-keV linking transition, but has no coincidence with the 339.3-keV transition.
So the 210.7-keV $\gamma$-ray is placed parallel to the 339.3-keV transition.

As to the position of the 362.3-keV transition, our suggestion is different from that of Ref.~\cite{Negi2012}. The 362.3-keV transition was placed parallel to the 339.3-keV transition in Ref.~\cite{Negi2012}.
As shown in Fig.~\ref{FIG:1103}(a), the $\gamma$-ray coincidence spectrum gated on the 362.3-keV transition, it is distinct that there is no coincidence between the 362.3-keV and the 575.6-keV transitions.
However, all the $\gamma$-rays decay from the levels below the 23/2$^{-}$ level at 3461.5 keV such as transitions with energies of 339.3-, 164.6-, 336.4-keV can be seen.
Therefore, the 362.3-keV transition is assigned as an intraband transition from 27/2$^{-}$ to 25/2$^{-}$, being parallel to the 575.6-keV transition.
In Fig.~\ref{FIG:1103}(b), the spectrum gated on the 1103-keV transition, there is a peak at energy of 1103 keV.
Therefore, the doublet $\gamma$-rays of 1103.3 keV is placed as a linking transition between 25/2$^{-}$ level at 4564.8 keV and 23/2$^{-}$ level at 3461.5 keV, which further confirms the position of the 362.3-keV transition.

The occurrences of the doublets of 493-keV and the 497-keV $\gamma$-transitions of bands 5 and 6 complicated the determination of the level scheme.
Note that the transitions with energies of 493 keV and 497 keV of bands 5 and 6 are doublet.
Both the 493- and 497-keV transitions can be seen in the spectra gated on their own energy, which were shown in Fig.~\ref{FIG:1103}(c) and Fig.~\ref{FIG:1103}(d), respectively.
The intensity profile of band 5 and band 6 also deserves attention. The 493- and 497-keV transitions are more intense than other transitions of bands 5 and 6.
Taking the spectrum gated on the 362.3 keV transition as an example, as shown in Fig.~\ref{FIG:1103}(a), the intensities of 493- and 497-keV transitions are almost twice of other $\gamma$-rays' of bands 5 and 6, such as the 515.0- and 367.0-keV transitions.
Evidence above clearly indicates that there are two $\gamma$-rays with almost the same energy of 493 and 497 keV in band 5 and band 6.
Besides, all the $\gamma$-rays of bands 5 and 6 and the linking transitions of 353.8 and 580.8 keV have mutual coincidence with the 497-keV transitions, while the 493-keV transitions have mutual coincident relationship with the 353.8-keV transition and all the intraband $\gamma$-rays of bands 5 and 6 except for the 266.0- and 243.0-keV transitions.
Moreover, the crossover transitions with energies of 1008.0 and 990.1 keV confirm the sequence of the transitions of 515.0, 493.0 and 497.1 keV.
Therefore, the doublets of 493keV are assigned in band 5 as 35/2$^{-}$$\rightarrow$33/2$^{-}$ and 31/2$^{-}$$\rightarrow$29/2$^{-}$, respectively, while the doublets of 497 keV transitions in band 5 as 29/2$^{-}$$\rightarrow$27/2$^{-}$ and band 6 as 39/2$^{-}$$\rightarrow$37/2$^{-}$, respectively.

The 990.1-keV crossover transition deserves more explanations, being different from Ref.~\cite{Negi2012}.
The 990.1-keV transition has mutual coincidence with the 824.7-keV transition, but it has no coincidence with the 1008.0-keV $\gamma$-ray. Such coincident relationship along with the energy balance restricts the position of the 990.1-keV $\gamma$-ray to 31/2$^{-}$ $\rightarrow$ 27/2$^{-}$ of band 5.

The newly added $\gamma$-rays with energies of 515.0, 581.5, and 685.1 keV have mutual coincidence with all the transitions of band 5 and extend this band to (39/2$^{-}$).
It is noted that there is a doublet peak around 581 keV, which could be distinguished from the spectra gated on the transitions of band 5 and band 6.
For example, a peak with the centroid energy of 581.5 keV is clearly seen in the spectrum gated on 493 keV, while a transition with centroid energy of 580.8 keV appears in the spectrum gated on 367.0 keV transition of band 6, as shown in Fig.~\ref{FIG:1103}(c) and Fig.~\ref{FIG:1103}(e), respectively.
This supports the assignment that $\gamma$-ray with energy of 581.5 keV is an intraband transition of band 5 while the 580.8-keV $\gamma$-ray is the interband transition linking 31/2$^{-}$ level of band 6 and 29/2$^{-}$ level of band 5.
The sequence of the newly identified $\gamma$-rays, i.e., 581.5-, 685.1- and the second 493.0-keV transitions are determined as shown in Fig.~\ref{FIG:109In} according to the intensities.

Band 6 is a new band built upon the 29/2$^{-}$ level at 5762.0 keV.
The $\gamma$-rays with energies of 353.8 and 580.8 keV link this band with band 5.
All the transitions of this band can be observed in Fig.~\ref{FIG:1103}(e).
The 243.0-keV $\gamma$-ray is placed based on the fact that it has mutual coincidence with all the $\gamma$-rays of band 6.
The 266.0-keV $\gamma$-ray has no coincidence with transitions which decay from the levels above 29/2$^{-}$ state at 5424.2 keV of band 5.
The existence of 353.8- and 580.8-keV transitions can also servers as a cross-check of the position of 266.0-keV $\gamma$-ray.
The sequence of other transitions of band 6 is mainly determined by the intensities.


\subsection{Band 7}\label{sec:band7}

Band 7, consisting of six $\triangle I$=2 transitions as 443.6, 654.0, 864.8, 829.6, 888.2 and 1000.4 keV, is built upon the 21/2$^{+}$ state of 4299.2 keV.
This band feeds to the 17/2$^{+}$ state at 2102.0 keV via the 1304.2- and 893.0-keV transitions.
Besides, it also decays to the state of 27/2$^{-}$ at 4927.1 keV of band 5 via the 469.7-keV transition.
The 864.8-keV transition together with 1304.2-, 893.0-, and 673.7-keV transitions has been reported in Ref.~\cite{Negi2012}.
Based on the multipolarity and newly found coincidence in the present work, we modify the placements of those transitions as band 7.

From Fig.~\ref{FIG:443}(a), a spectrum gated on the 864.8-keV transition, the 469.7-, 1304.2-, 893-keV linking transitions and all the members of band 7 can be clearly seen.
The insert of Fig.~\ref{FIG:443}(a) also shows the coincidence with the characteristic $\gamma$-rays of band 5 and band 1.
The transition with the energy of 469.7 keV plays a major role to determined the sequence of those $\gamma$-rays.
From the spectrum gated on the 469.7 keV, shown in Fig.~\ref{FIG:443}(b), the 362.3-keV transition along with those transitions in the lower part of band 1 can be seen.
Moreover, the 469.7-keV transition has mutual coincidence with the 864.8-, 829.6-, 888.2- and 1000.4-keV transitions of band 7.
However, transitions with energies of 443.6, 654.0, 893 and 1304.2 keV have no coincidences with the 469.7-, 362.3-keV $\gamma$-rays and transitions in the lower part of band 1.
The spectrum gated on the 443.6-keV transition is shown in Fig.~\ref{FIG:443}(c) as an example.

The multipolarity is also a main basis to establish band 7.
The DCO ratios of 893.0 and 1304.2 keV transitions extracted from the spectrum gated on the 673.7 keV $E2$ transition are around 0.5, which corresponds to a $\triangle I$=1 transition according to the DCO measurements of our experiments. Meanwhile the DCO ratios of the $\gamma$-rays of band 7 extracted in the same way are all around 1 which corresponds to a $\triangle I$=2 transition. So that band 7 is supposed to be composed of a sequence of $E2$ transitions.

It is noted that the intensity of 1304.2-keV $\gamma$-ray is particularly weaker than those of 443.6- and 893.0-keV transitions.
One possible explanation is that the flux from the 4299.2 keV state is divided into several weaker transitions. However, no other discrete transitions from this level have been found.


\subsection{Band 8}\label{sec:band 8}

Band 8 is observed for the first time.
The sum spectrum gated on the 941.6 and 816.0 keV transitions, shown in Fig.~\ref{FIG:443}(d), displays all the peaks of this band.
Those transitions above the 27/2$^{+}$ level at 5218.7 keV have mutual coincidence with the 475.9- and 443.6-keV transitions.
However, the decay path for this band to the lower states is still not clear.
Except for the 941.6-, 837.0- and 1099.6-keV transitions, others transitions of band 8 have coincidences with the 1428.3-, 402.0-, 1026.3-keV characteristic $\gamma$-rays of $^{109}$In.
A possible unobserved transition path from 19/2$^{+}$ at 4097.1 keV to the 13/2$^{+}$ at 1428.3 keV is indicated in Fig.~\ref{FIG:109In}.


\section{DISCUSSTION}\label{sec:discusstion}

\subsection{Configuration assignment}\label{sec:config}

\begin{figure}
  \includegraphics[width=10cm]{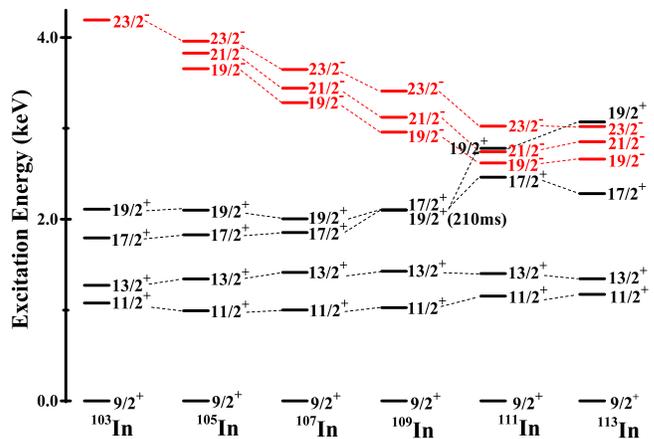}
  \caption{\label{FIG:level} (Color online) Systematics of the observed yrast states in the odd-$A$ Indium isotopes.}
\end{figure}

The low-lying structure of $^{109}$In can be described well in terms of a $g_{9/2}$ proton-hole coupled to excitations of the adjacent Sn core~\cite{Van1979,Kownacki1997}.
In Fig.~\ref{FIG:level}, the evolution of the yrast states in odd-$A$ indium isotopes~\cite{Kownacki1997,MR_107In,MR_111In,113In} is shown.
As Ref.~\cite{Kownacki1997} pointed out, the level spacing between 17/2$^{+}$ and 19/2$^{+}$ levels decreases with the increasing neutron number and almost reduces to zero in $^{109}$In, where the 19/2$^{+}$ is driven to be a spin-isomer.
As to the negative states, the excitation energy of 19/2$^{-}$ state decrease successively, which results from the fact that with the increasing neutron number the Fermi surface comes closer to the $h_{11/2}$ orbital.
The excitation energy of 19/2$^{-}$ state in $^{113}$In increases slightly probably due to fully occupying the $gd$ neutron subshell.
Similar negative-parity rotational bands built on the yrast 19/2$^{-}$ state can be found in $^{105, 107, 111}$In~\cite{Kownacki1997,MR_107In,MR_111In} and the configurations of those bands before the backbending in $^{105}$In and $^{107}$In were assigned as $\pi[g^{-1}_{9/2}]\bigotimes\nu[(g_{7/2}d_{5/2})^{1}(h_{11/2})^{1}]$ in Ref.~\cite{MR_107In}. Thus, the configuration of the rotational band built on the 19/2$^{-}$ state in $^{109}$In (band 1 in Fig.\ref{FIG:109In}) is suggested to be $\pi[g^{-1}_{9/2}]\bigotimes\nu[(g_{7/2}d_{5/2})^{1}(h_{11/2})^{1}]$.

\begin{figure}
  \includegraphics[width=8cm]{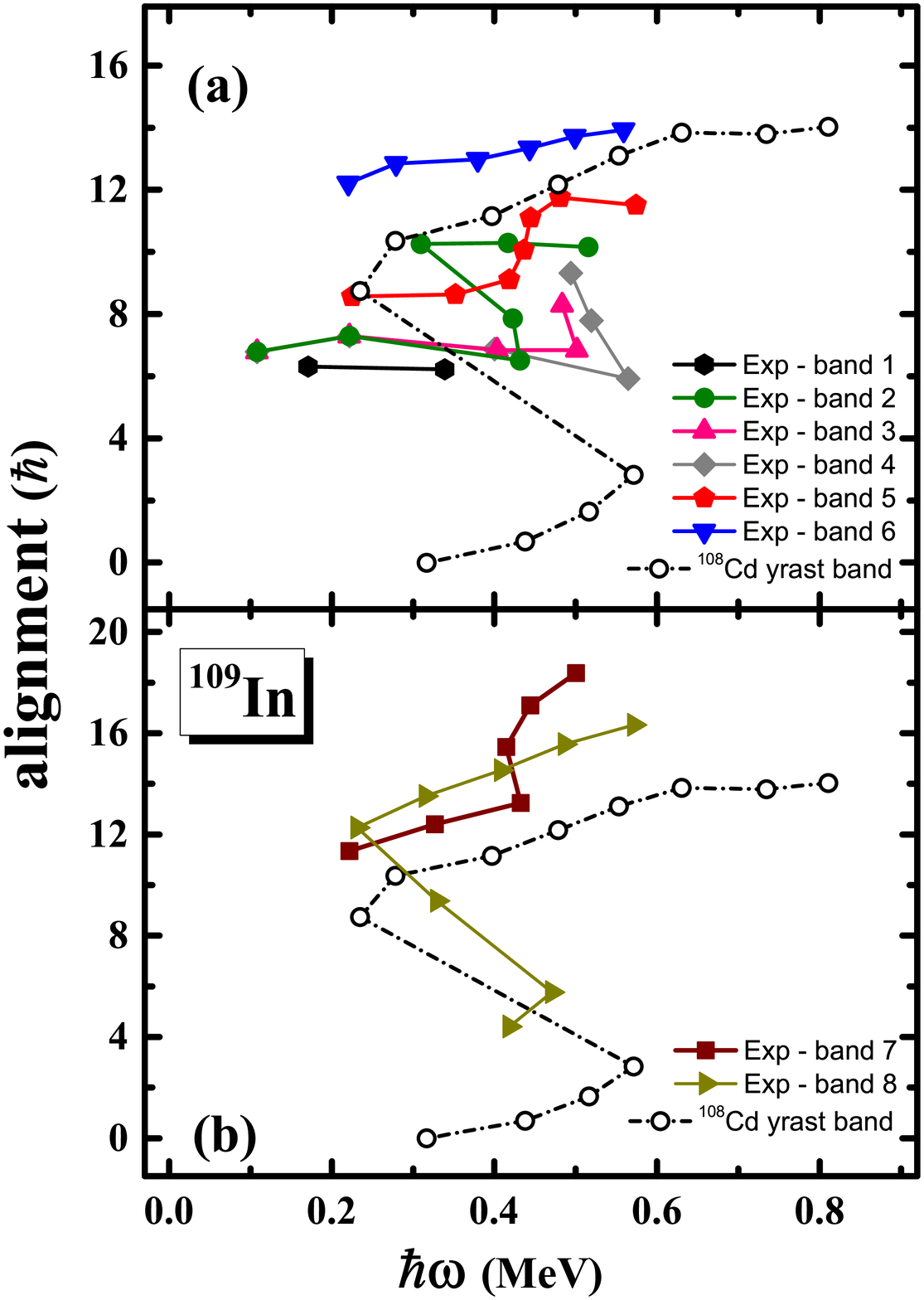}
  \caption{\label{FIG:ixw} (Color online) Experimental alignments as a function of rotational frequency $\hbar\omega$ for (a) bands 1-6 (b) bands 7, 8 in $^{109}$In and the yrast band in $^{108}$Cd~\cite{108Cd}, relative to a Harris parametrization of the core of $J_{0}=7\hbar^{2}$MeV$^{-2}$ and $J_{1}=9\hbar^{4}$MeV$^{-3}$.}
\end{figure}

In Fig.~\ref{FIG:ixw}, the experimental alignments of the bands in $^{109}$In are compared with that of the yrast band of the neighboring even-even nucleus $^{108}$Cd.
The alignment spin of band 1 is around 6$\hbar$ related to the yrast band of $^{108}$Cd, which is consistent with the configuration discussed above.
For band 2, a sharp backbending occurs at 0.42 MeV with a gain in aligned spin of about 3$\hbar$.
The alignments of this band before the backbending have almost the same variation trend as band 1 and are about 1.5$\hbar$ larger than that of band 1.
Moreover,  the bandhead energies of band 2 and band 1 are close.
It is reasonable to believe that the configuration of  band 2 before the backbending is similar to band 1.
Concerning of the mixture of $g_{7/2}$ and $d_{5/2}$ subshells, the difference between band 2 before the backbending and band 1 may be caused by the neutron occupying different $gd$ orbits, while the backbending of band 2 may be related to the alignment of neutrons in $gd$ orbitals.
Hence, we suggest the configuration of band 2 after the backbending as $\pi[g^{-1}_{9/2}]\bigotimes\nu[(g_{7/2}d_{5/2})^{3}(h_{11/2})^{1}]$.

Bands 3 and 4 are both connected to band 2 at the level of 25/2$^{-}$ with the energy of 3800.0 keV.
From Fig.~\ref{FIG:ixw}(a), backbending can be found for both bands 3 and 4 at close rotational frequencies.
Although the entire backbending has not been observed, the variations of experimental alignments for bands 3 and 4 are similar.
Therefore, those two bands are developed from $\pi[g^{-1}_{9/2}]\bigotimes\nu[(g_{7/2}d_{5/2})^{1}(h_{11/2})^{1}]$ and share the similar configuration.

\begin{figure}
  \includegraphics[width=8cm]{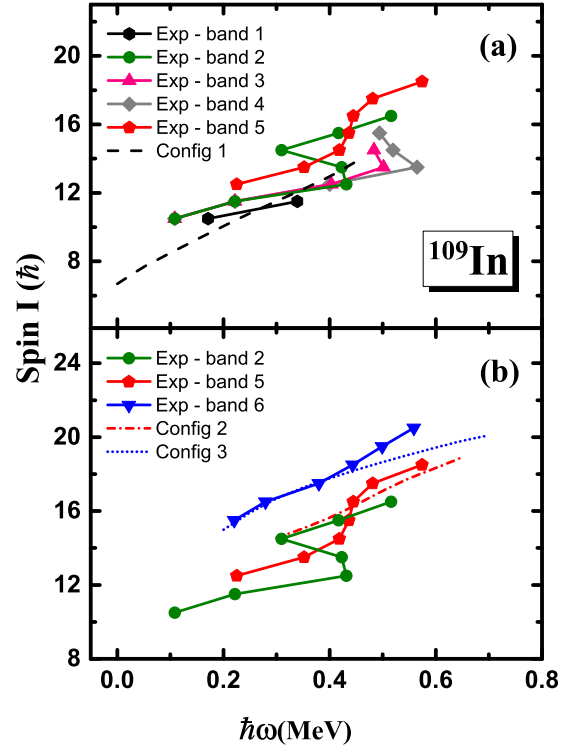}
  \caption{\label{Fig:TACIw} (Color online) Total angular momenta as a function of the rotational frequency in configuration-fixed (lines) constrained TAC-CDFT calculations with effective interaction PC-PK1~\cite{zhao2010new} compared with the data for (a) bands 1-5 (b) bands 2, 5, 6 in $^{109}$In.}
\end{figure}

Band 5 has a much gentle backbending at 0.42 MeV.
The alignments of this band before the backbending are approximately parallel to band 1, while the bandhead energy of band 5 is almost 1.5 MeV higher than that of band 1 as shown in Fig.\ref{FIG:109In}.
This can be explained by the unpaird neutrons in the $gd$ orbitals, i.e., one neutron is excited to $h_{11/2}$ orbital and the other is excited to higher-$\Omega$ $gd$ subshell.
As to the gentle backbending of band 5, it may be caused by one more pair of neutrons of $gd$ orbital being broken, which then contribute to the aligned spin.
The aligned spin of band 6 is around 12$\hbar$.
Although the alignments of band 6 are close to that of the yrast band of $^{108}$Cd after the backbending, which is due to the alignment of two $h_{11/2}$ neutrons, the trend of increment of the alignment of band 6 is much more gradual.
More importantly, the bandhead energy of band 6 is almost the same as the energy of the level at which the backbending is occurred in band 5. Thus, the configuration of band 6 can be suggested as $\pi[g^{-1}_{9/2}]\bigotimes\nu[(g_{7/2}d_{5/2})^{3}(h_{11/2})^{1}]$.

\begin{table*}[htbp]
\caption{\label{Table:config}The deformation parameters $\beta_{2}$ and $\gamma$ , and their corresponding unpaired nucleon configurations as well as the parities for configurations Config1-3 in the TAC-RMF calculations. Note that the valence nucleon configuration of Config1-3 is the same as $\pi[g^{-1}_{9/2}]\bigotimes\nu[(g_{7/2}d_{5/2})^{9}(h_{11/2})^{1}]$.}
\scalebox{1.1}{
\setlength{\tabcolsep}{22pt}
\begin{tabular}{cccc}
\hline
\hline
Notation  &  ($\beta$$_{2}$,$\gamma$) &  Unpaired nucleon configuration & $\pi$\\
\hline
Config1  &  (0.17,14.9$^{\circ}$)  &   $\pi[g^{-1}_{9/2}]\bigotimes\nu[(g_{7/2}d_{5/2})^{1}(h_{11/2})^{1}]$ & -\\
Config2  &  (0.14,11.4$^{\circ}$)  &  $\pi[g^{-1}_{9/2}]\bigotimes\nu[(g_{7/2}d_{5/2})^{3}(h_{11/2})^{1}]$ & -\\
Config3  &  (0.14,37.9$^{\circ}$)  &  $\pi[g^{-1}_{9/2}]\bigotimes\nu[(g_{7/2}d_{5/2})^{3}(h_{11/2})^{1}]$ & -\\
\hline
\end{tabular}}
\end{table*}

In Fig.~\ref{FIG:ixw}(b), bands 7 and 8 are compared with the yrast band of $^{108}$Cd.
A sharp backbending is evident at 0.43 MeV for band 8. The variation trend of alignments is approximately parallel to that of yrast band of $^{108}$Cd after the backbending, thus the $h_{11/2}$ neutrons is expected to be presented.
The aligned spin of band 8 is nearly 3.5$\hbar$ greater than that of the $^{108}$Cd, which can be caused by the occupation of the odd proton in the $g_{7/2}$ orbital.
Given the above speculations, band 8 is most likely based on $\pi$$g_{7/2}$$g_{9/2}$$^{-2}$ before the backbending and $\pi$$g_{7/2}$$g_{9/2}$$^{-2}\bigotimes$$\nu$h$_{11/2}$$^{2}$ after backbending.
For band 7, another band dominated by $\Delta$$I$=2 cascades, the initial aligned spin is around 11$\hbar$ and backbending occurs at 0.45 MeV.
The trend of alignments of band 7 before the backbending is parallel to band 8 after backbending and is 1$\hbar$ smaller than that.
Therefore, the configuration of band 7 should be $\pi$$d_{5/2}$$g_{9/2}$$^{-2}\bigotimes$$\nu$h$_{11/2}$$^{2}$ and the backbending might be attributed to the alignment of other $h_{11/2}$ neutrons.

\subsection{Theoretical interpretation}\label{sec:theoretical}

In the following, the structure of rotational bands in $^{109}$In will be investigated using the tilted axis cranking covariant density function theory (TAC-CDFT).
In contrast to its nonrelativistic counterparts~\cite{1}, the CDFT takes the fundamental Lorentz symmetry into account from the very beginning so that naturally takes care the important spin degree of freedom, resulting in great successes on many nuclear phenomena~\cite{1,2,3,4,5,6} and nucleosynthesis calculations~\cite{sun1,Sun2008Chin.Phys.Lett.2429,sun2,sun3,sun4,sun5}, such as well reproduction the isotopic shifts in Pb~\cite{7,8}, interpretation of the origin of pseudospin symmetry~\cite{9,10,11,12}, prediction of the spin symmetry in the anti-nucleon spectrum~\cite{13}, discovery of the multiple chiral doublet bands (M?D)~\cite{14,15}, as well as investigation of the charge-exchange excitations~\cite{16,17} and nuclear binding energy predictions~\cite{sun6,sun7,NIU201848}.
Particularly for describing nuclear rotations, it can include nuclear magnetism and then provides a consistent description of the currents and time-odd fields~\cite{18,19}.
The TAC-CDFT has been successfully used to describe magnetic rotation bands~\cite{024313,zhao2010new,yu2012}, anti-magnetic rotation bands~\cite{zhao2011prl,zhao2012prc}, chiral doublet bands~\cite{zhao2017plb}, linear alpha cluster bands~\cite{zhao2015prl}, and transitions of nuclear spin orientation~\cite{zhao2015prc}.

In the present TAC-CDFT calculations, the point-coupling interaction PC-PK1~\cite{zhao2010new} is adopted and the pairing correlations are neglected. The Dirac equation for the nucleons is solved in a spherical harmonic oscillator basis in Cartesian coordinates with $N_f=10$ major shells.
Aiming to describe the negative bands observed in $^{109}$In, three possible configurations are obtained and they correspond to a same valence nucleon configuration $\pi[g^{-1}_{9/2}]\bigotimes\nu[(g_{7/2}d_{5/2})^{9}(h_{11/2})^{1}]$.
The corresponding unpaired nucleon configurations with their deformation parameters and parities are listed in Table \ref{Table:config}.
Config1 is a three quasiparticle configuration, whereas both Config2 and Config3 are five quasiparticle configurations.
As the strong mixing between the $g_{7/2}$ and $d_{5/2}$ orbitals caused by the quadrupole and triaxial deformation in $^{109}$In, it is difficult to distinguish those orbitals in the theoretical calculations.
Accordingly, the unpaired nucleon configurations of both Config2 and 3 are written the same as $\pi[g^{-1}_{9/2}]\bigotimes\nu[(g_{7/2}d_{5/2})^{3}(h_{11/2})^{1}]$ in Table \ref{Table:config}, although their neutrons may actually occupy the different $gd$ orbits.

In Fig.~\ref{Fig:TACIw}, the calculated total angular momenta for each configuration are also shown as functions of the rotational frequency in compared with the experimental values of bands 1-6.
As shown in Fig.~\ref{Fig:TACIw}(a), the calculated results based on Config1 can be approximated to describe the experimental data for band 1.
The difference between calculated and experimental spectra may result from the neglecting of pairing correlations in the present calculation.
Band 2 and band 5 before backbending show the similar patterns as band 1, it is expected that their intrinsic configurations could be very similar, at least with the same high-$j$ orbital occupation.
It can be seen that in Fig.~\ref{Fig:TACIw}(b), the calculated results based on Config2 are in good agreements with data for bands 3 and 5 after back-bending, while the Config3 are in good agreements with data for band 6.
The configurations predicted here are consistent with the suggested configurations from the systematics discussed in Section \ref{sec:config}.

\subsubsection{$\gamma$ deformation and shape evolution}\label{sec:deformation}

\begin{figure*}[htbp]
  \includegraphics[width=13cm]{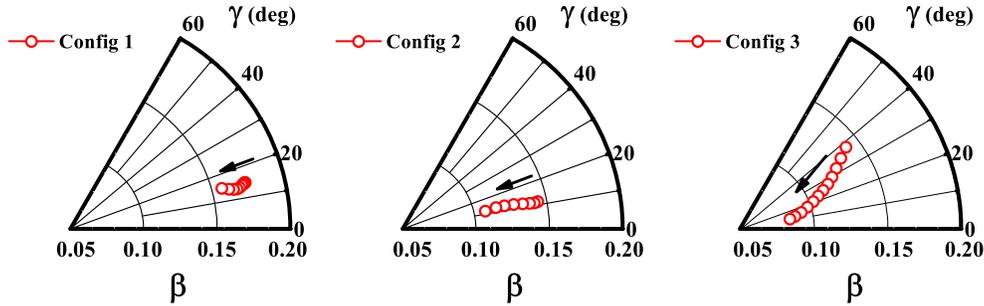}
  \caption{\label{Fig:deform} (Color online) Evolutions of deformation parameters $\beta$ and $\gamma$ driven by increasing rotational
frequency in the TAC-CDFT calculations for Config~1-3 in $^{109}$In. The arrows indicate the increasing direction of rotational frequency.}
\end{figure*}

\begin{figure*}[htbp]
  \includegraphics[width=13cm]{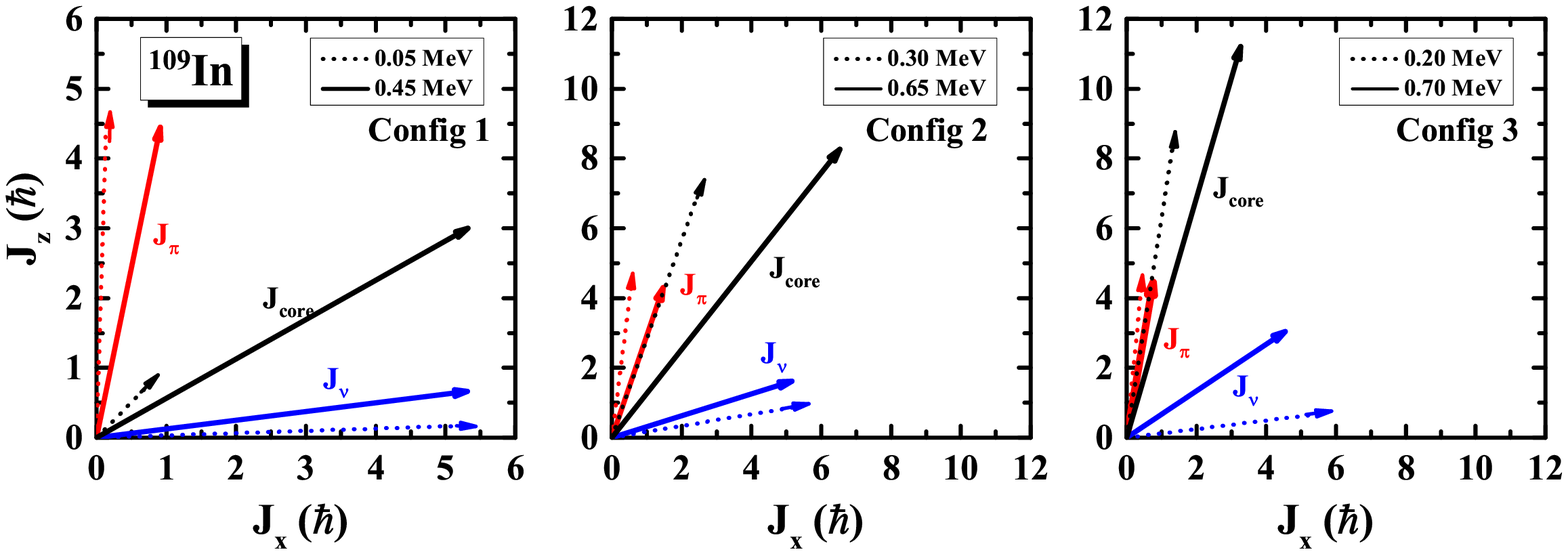}
  \caption{\label{Fig:vector} (Color online) The proton, neutron and core angular momentum vectors ($\bm{J}_{\pi}$, $\bm{J}_{\nu}$ and $\bm{J}_{core}$ ) for Config 1-3 in $^{109}$In at both the minimum and the maximum rotational frequencies in the TAC-RMF calculations.}
\end{figure*}

For $A$$\sim$110 mass region, the proton and neutron Fermi levels lie near different intruder orbitals. Equilibrium deformation of the nucleus will depend on the interplay of different driving forces when two or more high-$j$ quasiparticles are involved.

The evolution of deformation parameters $\beta$ and $\gamma$ obtained in the TAC-CDFT calculations are shown for Config1-3 in Fig.~\ref{Fig:deform}. The arrows indicate the direction towards increasing rotational frequency.
For Config1,  the $\gamma$ value stays around 15$^{\circ}$ and the $\beta$ value has a minor change with the rotational frequency.
The $\beta$ value of Config2 decreases from 0.143 to 0.107 with the increasing rotational frequency, while  the $\gamma$ value is nearly constant.
The deformation of Config3 has rather intense change with the rotational frequency, which the $\beta$ value decreases from 0.141 to 0.088 and the $\gamma$ value rapidly decreases from 37.94$^{\circ}$ to 12.9$^{\circ}$.
To sum up, a stable triaxial deformation is indicated for Config1, while a shape evolution of Config2 and Config3 can be clearly seen.
Especially for Config3, which is found to be fairly soft with respect to the $\gamma$ and $\beta$ deformation, it is predicted to suffer an evolution from a triaxial shape to a nearly spherical shape.

\subsubsection{Tilted rotation and possible chirality}\label{sec:chirality}

The mixture of $gd$ orbitals makes it intricacy to investigate the role of individual valence nucleons. For a concise physical image, the neutrons in $gd$ orbitals are counted into the core in the following discussion.
Figure~\ref{Fig:vector} illustrates how the total angular momentum $\overrightarrow{I}$ is generated.
The angular-momentum of the $g_{9/2}$ proton ($\overrightarrow{\bm{J}_\pi}$) and the $h_{11/2}$ neutron ($\overrightarrow{\bm{J}_\nu}$) as well as the core ($\overrightarrow{\bm{J}_{core}}$) at both the minimum and the maximum rotational frequencies in the TAC-CDFT calculations are shown in Fig.~\ref{Fig:vector}.
For the bands built on Config1-3, at low frequency, $\overrightarrow{\bm{J}_\nu}$ is along the short axis because of the neutron particle filling the bottom of the $h_{11/2}$, and the $\overrightarrow{\bm{J}_\pi}$ along the long axis because of the proton hole at the upper end of the $g_{9/2}$ shell.
As the rotation frequency increases, the angular momenta of the $g_{9/2}$ proton and the $h_{11/2}$ neutron come close to each other and contribute larger total spin, corresponding to the shears mechanism.
Furthermore, as the rotation frequency increases, the angular momentum of the core also increases remarkably, which mainly comes from the contribution from the neutrons in $gd$ orbital and shows the considerable role of collective rotation.
Consequently, the titled rotation is evident for three configurations of $^{109}$In.

Although the current two-dimensional TAC-CDFT theoretical calculation can not describe the non-planar rotation, the high-$j$ particle hole configuration and triaxial deformation indicate that the possibility of chirality should be taken into consideration.

\begin{figure}[htbp]
  \includegraphics[width=9cm]{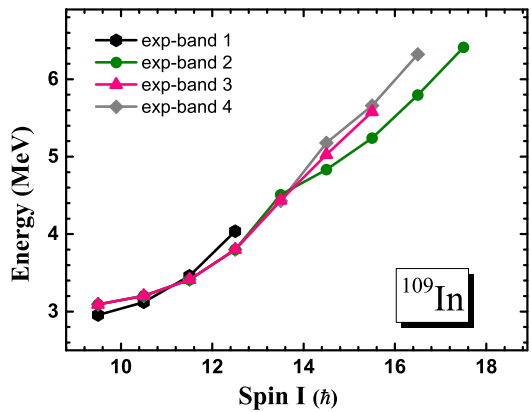}
  \caption{\label{Fig:EI} (Color online) Excitation energies as a function of the spin for the rotational bands in $^{109}$In.}
\end{figure}

The observation of two near-degenerate $\Delta I=1$ bands has been considered as the fingerprint of the existence of the chiral rotation.
The excitation energies of bands 1-4 are shown in Fig.~\ref{Fig:EI}.
The curves of band 1 and band 2 maintain a roughly constant energy difference of $\sim$100 keV, however, a crossover of the excitation energy occurs at 11.5$\hbar$.
Furthermore, as mentioned in Section.~\ref{sec:config}, despite the similar trends of the alignments for band 1 and band 2, the values of the alignments are not the same, which indicates that those bands may based on different intrinsic structure.
Based on the points discussed above, band 1 and band 2 are not chiral bands.
Band 3 and band 4 share the same decay path. The observed backbending pattern indicates they may be developed on the same configuration, as mentioned in Section.~\ref{sec:config}.
Moreover, as shown in Fig.~\ref{Fig:EI}, the excited energies of bands 3 and 4 are near-degenerate.
Thus bands 3 and 4 may be candidates for chiral bands. Further experimental data in $^{109}$In are called for to clarify such possibility.

\section{SUMMARY}

In summary, the high-lying states of $^{109}$In were investigated using the in-beam $\gamma$-ray spectroscopic techniques with the $^{100}$Mo($^{14}$N, 5$n$)$^{109}$In fusion-evaporation reaction.
The previously reported level scheme of $^{109}$In was extended considerably by adding 46 new $\gamma$ rays and establishing 5 new bands.
The configurations were tentatively assigned with the help of the systematics of neighboring odd-$A$ indium isotopes and the experimental aligned angular momenta.

The dipole bands in $^{109}$In are proposed to be based on one proton hole at the $g_{9/2}$ orbital and two or four unpaired neutrons at $g_{7/2}$, $d_{5/2}$ and $h_{11/2}$ orbitals and are compared with the titled axis cranking calculation in the framework of covariant density function theory.
The results shows that the shape of $^{109}$In undergoes an evolution on both $\beta$ and $\gamma$ deformations and a possible chirality is suggested in $^{109}$In.
Further experimental and theoretical efforts on this nucleus are crucial to verify this possibility.
\\

\acknowledgments The authors thank the crew of the HI-13 tandem accelerator at the China Institute of Atomic Energy for their help in steady operation of the accelerator and for preparing the target and S.~Q. Zhang and J. Meng for stimulating discussions, pre-reading and constructive comments on the manuscript.
This work is partially supported by the National Natural Science Foundation of China under Contracts No. 10975191, No. 11375023, 11475014 and No. 11575018, and  by the National Key R\&D program of China (2016YFA0400504).

\bibliography{mybib} 
\end{document}